\documentclass[twocolumn,showpacs,showkeys,preprintnumbers,amsmath,amssymb]{revtex4}
\usepackage{cases}
\usepackage{amssymb}
\usepackage{txfonts}
\usepackage{graphicx}
\usepackage{dcolumn}
\usepackage{bm}
\usepackage{amssymb}
\usepackage{amsmath}
\usepackage{epsfig}
\usepackage{subfigure}
\usepackage{overpic}
\begin{document}
\title{Pumping-assisted multistability of exciton-polariton condensates}
\author{Zi-Fa Yu$^{1,2}$}
\author{Ju-Kui Xue$^{2}$}
\author{Lin Zhuang$^{3}$}
\author{Jinkui Zhao$^{1,4}$}
\author{Wu-Ming Liu$^{1,4}$}
\email[Corresponding author. Email: ]{wliu@iphy.ac.cn}
\address{$^{1}$Beijing National Laboratory for Condensed Matter Physics, Institute of Physics, Chinese Academy of Sciences, Beijing 100190, China}
\address{$^{2}$Key Laboratory of Atomic $\&$ Molecular Physics and Functional Materials of Gansu Province, College of Physics and Electronic Engineering, Northwest Normal University, Lanzhou, 730070, China}
\address{$^{3}$State Key Laboratory of Optoelectronic Materials and Technologies, School of Physics, Sun Yat-Sen University, Guangzhou 510275, China}
\address{$^{4}$Songshan Lake Materials Laboratory, Dongguan 523808, China.}
\begin{abstract}
We investigate the multistability of exciton-polariton condensates excited by a nonresonant pump. An increase in pumping power moves the system away from non-Hermitian spectral degeneracy towards spectrum splitting through an exceptional point, which induces a transition from monostability to multistability. In the region of multistability, the system contains one steady and two metastable states. The analyses of stability show that metastable states maintain a finite lifetime and eventually evolve to steady states. A steady state with multi-peak soliton  different from general single-peak soliton is discovered for attractive polariton-polariton interaction. Moreover, we depict the diagram of the multistability in full parameter space to accurately manipulate the multistability. Our results open up exciting possibilities for controlling non-Hermitian quantum multistable states, which may be useful to designing polariton-based devices exploiting optical multistability.
\end{abstract}
 \maketitle

Exciton-polariton Bose-Einstein condensates (EPCs) in quantum-well semiconductor microcavities have provided a new platform for classical \cite{copy1} and quantum \cite{copy2} simulators, owing to room-temperature condensation \cite{copy3,copy4,copy5,copy6,copy7}, direct momentum and real-space imaging through the cavity photoluminescence \cite{copy8,copy9}, quantum nonequilibrium and non-Hermitian nature \cite{copy10,copy11,copy12,copy13}, and highly controllability via suitably manipulating both optical pump and quantum-well microcavities \cite{copy14}. In contrast to traditional condensates, the nonequilibrium and non-Hermitian nature of EPCs requires an external continuous pump to maintain the population owing to their short lifetime induced by radiative decay \cite{copy10,copy11,copy12,copy13}. Consequently, EPCs are truly dynamical steady states instead of thermal equilibrium states. Such self-localized steady states are spontaneously created even without any external trap potential. They have been widely discussed in theory and experiment \cite{copy15,copy16,copy17,copy18,copy19,copy20,copy21,copy22,copy23,copy24,copy25,copy26,copy27,copy28},
including solitons \cite{copy18,copy19,copy20,copy21,copy22,copy23}, vortices \cite{copy24,copy25}, vortex lattices \cite{copy26,copy27}, and vortex pairs \cite{copy28}. However, there are still some novel properties of steady states worth exploring, and they have no analogues in thermal equilibrium states.

Recently, a nontrivial phenomenon of steady states, i.e., optical multistability, has been reported in EPCs  \cite{copy29,copy30,copy31,copy32,copy321,copy322}, where the system contains multiple stable states for a given set of parameters. It has a potential application in optical circuits, optical computing, and all-optical switches \cite{switch1,switch2,switch3}. It has been explored in different nonlinear optical system \cite{copy33,copy34,copy343,copy344}, such as optical fibers \cite{copy343}, and photonic crystals  \cite{copy344}. Vortex multistability with different topological
charges is discovered for the same system and excitation parameters, and is used to manipulate vortex multiplets \cite{copy29}. Parity bifurcation transition of multistable states appears in organized phase-locked EPCs \cite{copy321}, and it comes from modulation instability \cite{copy322}. However, the corresponding non-Hermitian spectrum of multistable states is still not clear, which can explain the appearance of multistability from another perspective. The spectral degeneracy structure of EPCs is modified by the non-Hermiticity, which affects the steady state, localization, nonlinearity, transportation, and dynamics of the system \cite{copy35,copy36,copy37,copy38,copy39}. In a non-Hermitian system, exceptional points \cite{copy40,copy41} (where multiple eigenstates collapse and corresponding eigenmodes coalesce into one) can cause a range of peculiar phenomena, such as unidirectional transmission \cite{copy42}, anomalous absorption \cite{copy43}, and chiral modes \cite{copy44}. It is important to understand multistability of EPCs via non-Hermitian spectrum, which may reveal the existence of new and novel quantum states.

\begin{figure}[htpb]
\begin{center}
\includegraphics[width=8.0cm]{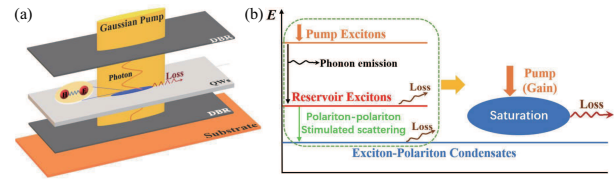}
\end{center}
\caption{(a) Schematic diagram of semiconductor microcavity device supporting exciton-polariton condensates (EPCs) in quasi-one-dimensional semiconductor microwire. The device is comprised of quantum wells (QWs) placed between two distributed Bragg reflectors (DBRs). Polaritons are excited by a nonresonant Gaussian pump beams incident from above. (b) Schematic representation of the mechanism for generating EPCs.}\label{fig7}
\end{figure}

In this Letter, we reveal the mechanism for the formulation of multistable states associated with the non-Hermitian spectrum splitting in EPCs excited by a nonresonant Gaussian pump (see Fig. \ref{fig7} (a)). In experiments \cite{copy3,copy4,copy5,copy6,copy7,copy8,copy9}, QWs are GaAs thin layers of the order of $10{\rm nm}$, and DBRs consist of multiple pairs of alternated AlAs and GaAs layers. EPCs have typical lifetimes of the order of $100 {\rm ps}$ and consideration temperature of the order of $10 {\rm K}$ in GaAs and CdTe semiconductors. The pump produces a reservoir of high energy excitons which scatter continuously into lower energy polaritons. When scattering amplification overcomes losses, condensates are formed. In the adiabatic approximation, the reservoir can be regarded as static and it moderates condensate densities. It can be modeled by considering an effective gain or pump, loss, and gain saturation (see Fig. \ref{fig7}(b)). Our results indicate that the increasing of pumping power moves the system away from the non-Hermitian spectral degeneracy towards spectrum splitting through an exceptional point, which induces the multistability of EPCs. The diagram of the multistability is depicted in full parameter space. A steady state with multi-peak soliton is discovered for attractive EPCs.

\emph{EPCs system:}
Motivated by the relevant experiment of EPCs \cite{copy3,copy4,copy5,copy6,copy7,copy17}, we consider a nonequilibrium EPCs system excited by a nonresonant one-dimensional continuous wave pump with a Gaussian profile (see Fig. \ref{fig7}). Under the mean field approach, EPCs can be described by a dimensionless open-dissipative Gross-Pitaevskii equation, incorporating pump, loss and gain saturation \cite{copy10,copy11,copy26,copy45,copy46,copy47}:
\begin{equation}\label{1}
{\rm i}\frac{\partial\psi}{\partial t}=\left[-\frac{1}{2}\nabla^{2}+g|\psi|^{2}+\frac{\rm i}{2}\left(P(x)-\gamma-\eta|\psi|^{2}\right)\right]\psi,
\end{equation}
where, the physical variables are rescaled as $\psi\thicksim l^{-1/2}\psi$, $t\thicksim\omega^{-1}t$, and $x\thicksim lx$ with condensate characteristic length $l=\sqrt{\hbar/(m\omega)}$ and characteristic frequency $\omega$. $m$ is the polariton effective mass. $g$ is the dimensionless polariton-polariton interaction constant. $P(x)$ is a spatially modulated Gaussian pump with power $P_{0}$ and width $w$, i.e., $P(x)=P_{0}{\rm e}^{-x^{2}/w^{2}}$. $\gamma$ is the polariton loss rate. $\eta$ refers to the gain saturation. For a nonresonantly pumped system, gain saturation is necessarily present. Without gain saturation, the condensate grows indefinitely when pump exceeds loss, and vanishes when pump falls short of loss \cite{copy26}. The steady state described by this saturation is similar to the one described by considering a static reservoir of noncondensed polaritons \cite{copy46}. The related parameters can be estimated by experiment \cite{copy17}: $m\sim 5\times10^{-5}m_{\rm e}$ with $m_{\rm e}$ being the free electron mass, $\omega\sim0.01{\rm ps}^{-1}$, $\gamma\sim\omega$, $g\sim7.9{\rm \mu eV\mu m^{2}}(\hbar\omega l^{2})^{-1}$, $\eta\sim g$. The system can be easily realized in the currently experimental conditions.

To acquire the steady-state solution and the dynamical evolution of the nonequilibrium system, we use the variational method for dissipative systems \cite{copy21,copy48}. A natural variational ansatz is the Gaussian trial distribution,
\begin{equation}\label{2}
\psi(x,t)\!=\!\left(\frac{n}{\sqrt{\pi}R}\right)^{1/2}\exp\left[\!-\!\frac{(x\!-\!\xi)^{2}}{2R^{2}}\!+\!{\rm i}k(x\!-\!\xi)\!+\!\frac{{\rm i}\delta}{2}(x\!-\!\xi)^{2}\!+\!{\rm i}\phi\right],
\end{equation}
which denotes that EPCs have a Gaussian distribution with the center mass position $\xi(t)$, momentum $k(t)$, size $R(t)$, related variation rate of width $\delta(t)$, phase $\phi(t)$, and the number $n(t)$ (i.e., $n=\int|\psi|^{2}{\rm d}x$) at a given time $t$. After variational analyses, we can acquire dynamical evolvement equations of related variational parameters $q_{i}(t)$ ($q_{i}=\{\xi,k,R,\delta,n,\phi\}$), stationary equations of revelent states (i.e., $q_{i}(t)\equiv\bar{q_{i}}$), and the non-Hermitian energy of the system (for details, please see Supplemental Material \cite{copy49}).

EPCs are formed when the pumping power is larger than the threshold $P_{0}^{\rm th}$, which can be obtained from the corresponding stationary equations \cite{copy49},
\begin{equation}\label{12}
P_{0}^{\rm th}=\sqrt{\frac{\gamma}{2}}\left(\frac{1+\sqrt{1+2\gamma w^{2}}}{w}\right).
\end{equation}
Obviously, the threshold $P_{0}^{\rm th}$ is independent of the polariton-polariton interaction and the gain saturation. It only depends on the pumping size and the polariton loss rate. For a homogeneous pump (i.e., $w\rightarrow\infty$), $P_{0}^{\rm th}=\gamma$. A nonresonant Gaussian pump can results in damped dipolar oscillation and create the self localization of EPCs \cite{copy49}.

\begin{figure}
\begin{center}
\includegraphics[width=8.0cm,height=6.0cm]{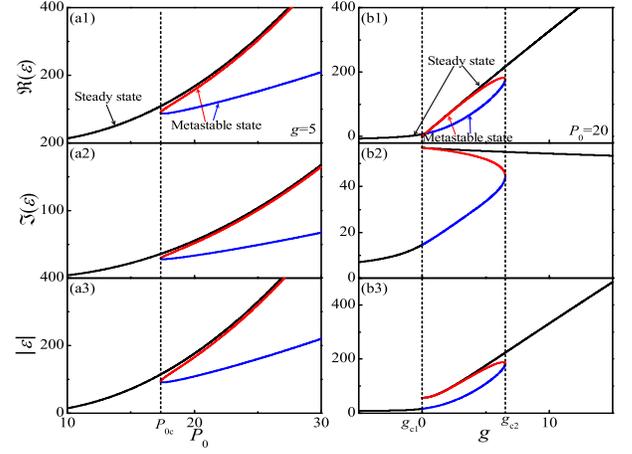}
\end{center}
\caption{Non-Hermitian spectrum of EPCs for different pumping power $P_{0}$ (a1)-(a3) and different polariton-polariton interaction $g$ (b1)-(b3) with $\gamma=5.0$, $\eta=5.0$, and $w=5.0$. Black solid lines correspond to steady state, while red and blue solid lines correspond to metastable state.}\label{fig6}
\end{figure}

\emph{Non-Hermitian spectrum:}
Figure \ref{fig6} demonstrates the non-Hermitian spectrum of EPCs for different pumping powers and polariton-polariton interactions. There is one exceptional point in the non-Hermitian spectrum for varying pumping power at $P_{0}=P_{\rm 0c}$ as shown in Figs. \ref{fig6}(a1)-\ref{fig6}(a3). For weak pumping powers, i.e., $P_{0}^{\rm th}<P_{0}<P_{\rm0c}$, the non-Hermitian spectrum is degenerate with only one energy, which corresponds to a steady state. For strong pumping powers, i.e., $P_{0}>P_{\rm0c}$, increasing pumping power moves the system away from the spectral degeneracy, and spectrum splitting occurs. In this case, there are three energies, where one is original spectral band corresponding to steady state (black lines), and the other two is emerging spectral bands corresponding to metastable states (red and blue lines). Thus, incoherent pumps can result in the spectrum splitting, which induces the emergence of multistability.

However, for varying polariton-polariton interactions, there are two exceptional points of non-Hermitian spectrum at $g=g_{\rm c1}$ and $g=g_{\rm c2}$ (Figs. \ref{fig6}(b1)-\ref{fig6}(b3)). For attractive interactions ($g<g_{\rm c1}$) and strong repulsive interactions ($g>g_{\rm c2}$), the spectrum is degenerate, which corresponds to a steady state. For weak repulsive interactions ($g_{\rm c1}<g<g_{\rm c2}$), the spectrum splitting leads to the emergence of metastable states (red and blue lines). The spectral degeneracy of polariton condensates is different from the case in atomic condensates. It is resulted from the non-Hermitian nature and is not caused by the nonlinearity like that in traditional condensates. Therefore, control of related parameters allows us to manipulate the approach to the exceptional point of the non-Hermitian spectrum and the generation of multistability for EPCs \cite{copy49}.

\begin{figure}
\begin{center}
\includegraphics[width=8.0cm,height=6.0cm]{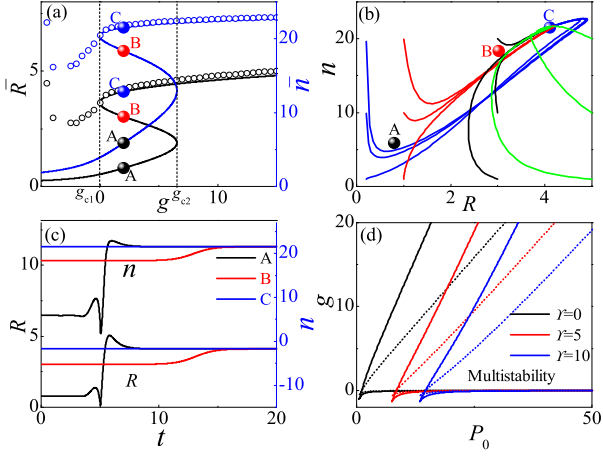}
\end{center}
\caption{Multistability of EPCs for $w=5$. (a) The EPCs size $\bar{R}$ and particle number $\bar{n}$ in multistable states as function of polariton-polariton interaction $g$ for $P_{0}=20$, $\gamma=5.0$,  and $\eta=5$ depicted by variational approach (solid lines) and numerical simulation of Eq. (\ref{1}) (circles). (b) Phase trajectory in $R-n$ plane for different initial state with $g=2.0$. (c) Temporal evolution of EPCs size $R$ and particle number $n$ with $g=2.0$ for different multistable states as marked A, B, and C in (a) and (b). (d) The diagram of multistability in $P_{0}-g$ plane with $\eta=5$ (short dotted lines) and $\eta=10$ (solid lines).}\label{fig2}
\end{figure}

\emph{Multistability:}
The multistability of EPCs is induced by strong nonresonant pump. It also depends on the polariton-polariton interaction, loss rate and gain saturation. The stability of the system can be explained by the corresponding dynamics of EPCs. We introduce damping coefficient $C_{B}$ for breathing dynamics and eigenvalues $\Lambda$ for the characteristic matrix of linearizing dynamic evolution equations \cite{copy49}. Then positive $C_{B}$ and negative $\Re(\Lambda)$ can indicate the existence of long-lifetime EPCs.

\begin{figure}
\begin{center}
\includegraphics[width=8.0cm,height=6.0cm]{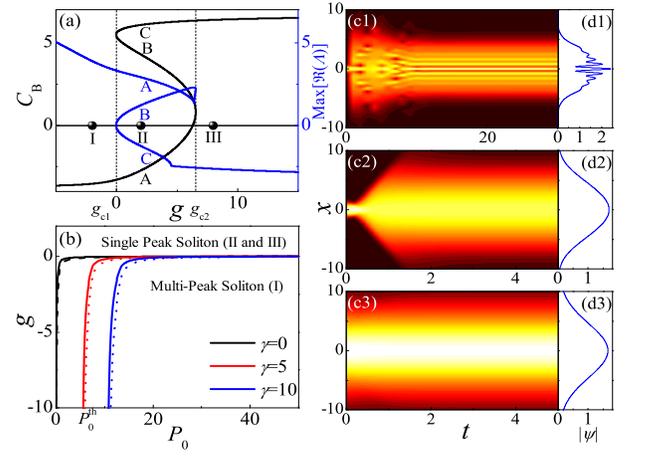}
\end{center}
\caption{The dynamics of EPCs with $w=5.0$. (a) $C_{B}$ and ${\rm Max}[\Re(\Lambda)]$ versus $g$ with $P_{0}=20$, $\gamma=5.0$ and $\eta=5.0$. (b) The diagram for the single- and multi-peak soliton in $P_{0}-g$ plane with $\eta=5.0$ (solid lines) and $\eta=10.0$ (dotted lines). (c1)-(c3) The temporal evolution of EPCs for $g=-2.0$, $2.0$, and $8.0$ as marked by I, II, and III in (a). (d1)-(d3) The final spatial distribution for EPCs at $t=50$ corresponding to (c1)-(c3).}\label{fig3}
\end{figure}

When $g_{\rm c1}<g<g_{\rm c2}$, the multistability of EPCs appears and the system exhibits three equilibrium states: two metastable states A and B and one steady state C (Fig. \ref{fig2}(a)). For state A, $C_{B}<0$ indicates the oscillation amplitude of EPCs size $R$ is increasing with time, while ${\rm Max[\Re(\Lambda)]}>0$ indicates that this state is unstable and can not be maintained for a long time (Fig. \ref{fig3}(a)). The driven oscillation of EPCs size leads to that this state eventually evolves into the steady state C with a larger size (see Figs. \ref{fig2}(c), \ref{fig3}(c2) and \ref{fig3}(d2)). State B is also maintained for a finite lifetime due to $C_{B}>0$ and ${\rm Max[\Re(\Lambda)]}>0$ (Fig. \ref{fig3}(a)). For state C, $C_{B}>0$ and ${\rm Max[\Re(\Lambda)]}<0$ indicate that it can exist for a long lifetime. In the region of the multistability, whatever the initial size and particle number of the polariton condensate are, the condensate will evolve into a steady state C. It may pass by a metastable state A or B in the evolutionary process, which depends on the initial state. This is depicted in the phase trajectory of the system as shown in Fig. \ref{fig2}(b). These states with a Gaussian distribution are also called single-peak solitons.

When $g>g_{\rm c2}$, there is only one steady state with single-peak soliton (Fig. \ref{fig2}(a)). $C_{B}>0$ and ${\rm Max[\Re(\Lambda)]}<0$ indicate that the system can eventually evolve into a steady state for arbitrary perturbation due to the damping effect. The steady state can still be maintained as a single-peak soliton (Figs. \ref{fig3}(c3) and \ref{fig3}(d3)). When $g<g_{\rm c1}$, EPCs only exhibits one equilibrium state (Fig. \ref{fig2}(a)). $C_{B}<0$ and ${\rm Max[\Re(\Lambda)]}>0$ indicate that this state is unstable (Figs. \ref{fig3}(a)). The driven oscillation of EPCs size leads to the breaking up of single-peak soliton, and eventually forming a multi-peak soliton as shown in Figs. \ref{fig3}(c1) and \ref{fig3}(d1). This results in the deviation between variational and numerical results in attractive interaction regions (Fig. \ref{fig2}(a)).

The generation of the multistability is also depends on other parameters, which can be obtained from the multistability diagram in the Fig. \ref{fig2}(d). For a given strong enough pumping power $P_{0}$, multistability emerges in the region of $g_{\rm c1}<g<g_{\rm c2}$, and outside this region, the system only exhibits one steady state. With the enhancement of $P_{0}$, $g_{\rm c1}$ first increases quickly then tends to zero, while $g_{\rm c2}$ increases linearly. Thus, the pumping power can enlarge the region of the multistability. On the other hand, the loss rate $\gamma$ causes the multistable region shift right, and the gain saturation $\eta$ causes $g_{\rm c2}$ shift up. Thus the loss rate (the gain saturation) can shrink (enlarge) the region of multistability.

The boundary between the single- and multi-peak soliton is demonstrated in Fig. \ref{fig3}(b). For $P_{0}>P_{0}^{\rm th}$, $g_{\rm c1}$ divides the diagram of the steady state into two, one is the multi-peak soliton region ($g<g_{\rm c1}$) and the other is the single-peak soliton region ($g>g_{\rm c1}$). As the enhancement of $P_{0}$, $g_{\rm c1}$ increases quickly and then tends to a constant less than zero. Thus, the generation of multi-peak soliton is induced by attractive polariton-polariton interaction. The loose rate and the gain saturation can both shrink the region of multi-peak solitons.

\begin{figure}
\begin{center}
\includegraphics[width=8.0cm,height=6.0cm]{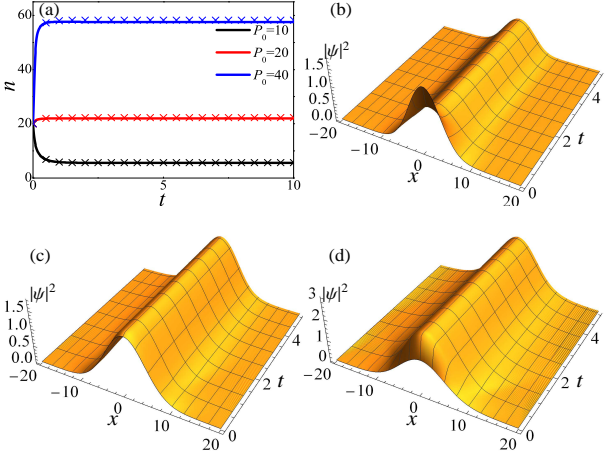}
\end{center}
\caption{Dissipative solitons for different pumping power $P_{0}$ with $g=\eta=\gamma=w=5.0$. (a) The temporal evolution of EPCs number depicted by analytical expression Eq. (\ref{16}) (solid lines) and numerical simulation of Eq. (\ref{1}) (crosses). (b)-(d) The corresponding temporal evolution of solitons depicted by numerical simulation of Eq. (\ref{1}) for the pumping power $P_{0}=10$, $20$, and $40$, respectively.}\label{fig4}
\end{figure}

\emph{Steady state:}
In the steady state, solving dynamic evolution equations \cite{copy49} results in the number of EPCs
\begin{equation}\label{16}
n(t)=\frac{\sqrt{2\pi}\bar{R}Gn_{0}}{\eta n_{0}+(\sqrt{2\pi}\bar{R}G-\eta n_{0})\exp(-Gt)},
\end{equation}
where, $n_{0}$ is initial polariton number, and $G=P_{0}w/\sqrt{w^{2}+\bar{R}^{2}}-\gamma$ is an effective net gain, which depends on the pumping power and size, the loss rate, and the EPCs size. For a homogeneous pumping (i.e., $w\rightarrow\infty$), the effective net gain $G=P_{0}-\gamma$. The EPCs number can also be depicted by the temporal evolution of the corresponding dissipative solitons, which is shown in Fig. \ref{fig4}. When EPCs are supersaturated initially, i.e., $n_{0}>\bar{n}$, where $\bar{n}=\sqrt{2\pi}\bar{R}G/\eta$ is the EPCs number in the steady state (also called the saturated state), it dissipates until $n=\bar{n}$ (see Figs. \ref{fig4}(a) and \ref{fig4}(b)). When EPCs do not reach to saturation, i.e., $n_{0}<\bar{n}$, the EPCs number quickly increases to $\bar{n}$ (see Figs. \ref{fig4}(a) and \ref{fig4}(d)). When EPCs are saturated, i.e., $n_{0}=\bar{n}$, the EPCs number still remains unchanged (see Figs. \ref{fig4}(a) and \ref{fig4}(c)). The saturated EPCs number $\bar{n}$ depends only on the effective net gain, the gain saturation, and the EPCs size. It is independent of the initial state of EPCs. It is clear that the temporal evolution of the
dissipative soliton is indeed in good agreement with our theoretical prediction of Eq. (\ref{16}).

\begin{figure}
\begin{center}
\includegraphics[width=8.0cm]{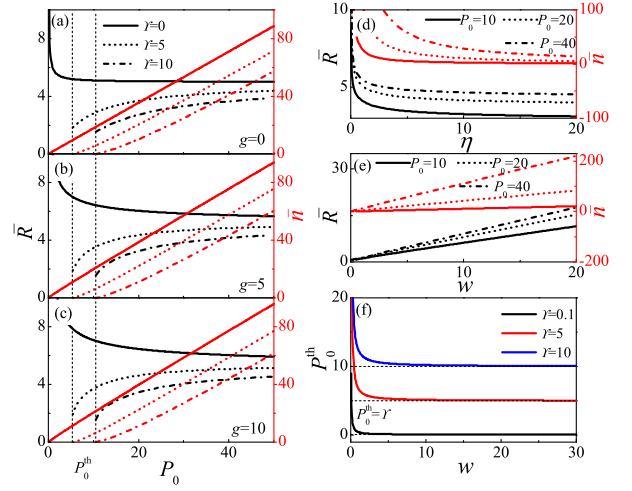}
\end{center}
\caption{Steady state of EPCs. (a)-(c) EPCs size $\bar{R}$ and number $\bar{n}$ as function of pumping power $P_{0}$ for different loss grate $\gamma$ and polariton-polariton interaction $g$ with pumping size $w=5$ and saturation rate $\eta=5$. (d)-(e) $\bar{R}$ and $\bar{n}$ as function of $\eta$ and $g$ for different $P_{0}$ with $\gamma=g=5$. (f) The critical pumping power for generating condensed polaritons $P_{0}^{\rm th}$ as function of $w$ for different $\gamma$.}\label{fig1}
\end{figure}

The size and particle number of EPCs in the steady state is clearly depicted in Fig. \ref{fig1}. As shown in Fig. \ref{fig1}(a), without polariton loss (i.e., $\gamma=0$), the increase of pumping power leads to that the EPCs size first quickly decreases then remains unchanged and the EPCs number always increases. In this case, the system is unstable because the EPCs size is larger than the pumping size and they will be diffuse. However, when considering the loss (i.e., $\gamma\neq0$) and $P_{0}>P_{0}^{\rm th}$, the EPCs size first quickly increases then tends to a constant and the EPCs number increases as pumping power enhances. When $P_{0}\leq P_{0}^{\rm th}$, there is no condensed polariton, i.e., $\bar{n}=0$. Furthermore, the polariton-polariton interaction $g$ and pumping size $w$ can increase the EPCs number and size, while the loss rate $\gamma$ and the gain saturation $\eta$ can decrease them. The critical pumping power threshold for generating condensed polariton $P_{0}^{\rm th}$ obtained by Eq. (\ref{12}) is reduced by the pumping size and promoted by the loss rate (see Fig. \ref{fig1}(f)).

In summary, we have demonstrated the multistability of EPCs in the region of non-Hermitian spectrum splitting by using variational analyses and direct numerical simulations. The multistable state manifests itself in one steady state and two metastable states, while two metastable states can only exist in a finite lifetime and eventually evolve into the steady state. We also construct the diagram of multistability and discover a steady state with multi-peak solitons in attractive EPCs. The mutistability and steady states of EPCs can be manipulated by approaching the exceptional point of non-Hermitian spectrum via appropriately adjusting related parameters. It could be useful in low-energy polariton-based devices exploiting optical multistability \cite{switch1,switch2,switch3}.

This work is supported by the National Key R$\&$D Program of China under Grants
No. 2016YFA0301500, NSFC under grants Nos. 11865014, 11764039, and 61835013, Strategic Priority Research Program of the Chinese Academy of Sciences under grants Nos. XDB01020300, XDB21030300, Scientific research project of Gansu higher education under Grand No. 2019A-014, the creation of science and technology of
Northwest Normal University, China, under Grants No. NWNU-LKQN-18-33.

\end{document}